\documentclass[%
 reprint,
 amsmath,amssymb,
 aps,
]{revtex4-2}

\usepackage{graphicx}
\usepackage{dcolumn}
\usepackage{bm}
\usepackage{xcolor}
\usepackage[normalem]{ulem}
\usepackage{siunitx}
\usepackage{float}
\usepackage{chngcntr}
\usepackage{verbatim}
\usepackage{tikz}
\newcommand*\circled[1]{\tikz[baseline=(char.base)]{
            \node[shape=circle,draw,inner sep=1pt] (char) {#1};}}

\begin{document}

\preprint{APS/123-QED}

\title{Intersonic Detachment Surface Waves in Elastomer Frictional Sliding}

\author{Huifeng Du$^{1}$}
\thanks{These authors contributed equally to this work.}
\author{Emmanuel Virot$^{2}$}
\thanks{These authors contributed equally to this work.}
\author{Liying Wang$^{3}$, Sam Kharchenko$^{3}$, \\Md Arifur Rahman$^{3}$, David A. Weitz$^{2}$}
\author{Shmuel M. Rubinstein$^{2}$} 
\author{Nicholas X. Fang$^{1}$}
\email{nicfang@mit.edu}

\affiliation{$^1$Department of Mechanical Engineering, Massachusetts Institute of Technology, 77 Massachusetts Avenue, Cambridge, Massachusetts 02139, USA \\ $^2$John A. Paulson School of Engineering and Applied Sciences,
Harvard University, Cambridge, Massachusetts 02138, USA \\ $^3$BASF Corporation, 1609 Biddle Avenue, Wyandotte, Michigan 48192, USA}










\begin{abstract}
Elastomeric materials when sliding on clean and rough surfaces generate wrinkles at the interface due to tangential stress gradients. These interfacial folds travel along the bottom of elastomer as surface detachment waves to facilitate the apparent sliding motion of elastomer. At very low sliding speed compared to elastic surface waves, the process is dominated by surface adhesion and relaxation effects, and the phenomenon is historically referred to as Schallamach waves. We report in this letter the observation of fast-traveling intersonic detachment waves exceeding the Rayleigh and shear wave velocities of the soft material in contact. The spatio-temporal analysis revealed the accelerating nature of the detachment wave, and the scaling of wave speed with the elastic modului of the material suggests that this process is governed by elasticity and inertia. 
Multiple wave signatures on the plot were connected to different stages of surface wrinkles, as they exhibited distinctive slopes (from which velocities were derived) in the generation, propagation and rebound phases. We also characterized the frequencies of wrinkle generation in addition to the speeds and found a consistent scaling law of these two wave characteristics as the stiffness of elastomer increased.
Physical implications of this new finding may further promote our understanding of elastomer noise generation mechanisms, as at macroscopic sliding velocity, the frequency of elastomer instability readily enters human audible ranges and interacts with other vibratory frequencies to cooperatively create harsh and detrimental noises in disc braking, wiper blade and shoe squeaking among many other elastomer applications.

\end{abstract}


\maketitle


\section{Introduction}
Elastomer exhibits distinctive behavior among many other materials when in sliding contact with hard substrates. While theoretical advancements in contact mechanics including the JKR \cite{johnson1971surface}, Maguis \cite{maugis1992adhesion} provide accurate predictions of normal contact configurations, tangential contact behaviors of elastomers deviate from conventional friction models dictated by the unique properties of elastomeric materials: relatively low moduli, viscoelastic behavior and internal friction \cite{persson1998theory} \cite{schallamach1971does}. Due to its high deformability, elastic instabilities are likely to form at the contact interface when compressive tangential forces are large enough to buckle the bottom surface, as were repeatedly observed and analyzed after Schallamach (1971) \cite{schallamach1971does} formally conceptualized the idea. The wrinkles generated on the buckled surface travel in the same direction as the bulk motion of the elastomer, as moving ridges carrying along tunnels of trapped air, forming 'waves of detachment' since elastomer surface loses contact with the substrate when lines of folds pass through. At low sliding velocities (e.g. less than \SI{500}{\micro\meter\per\second} \cite{rand2006insight}), a competition between adhesion forces at the interface and the relaxation effects governs the generation of surface instability \cite{rand2006insight} \cite{barquins1975rubber}. The interfacial adhesion strength exhibits strong dependence on velocity due to the rate-dependent bond dissociation kinetics (\cite{chaudhury1999rate}. At higher sliding velocities (typically around \SI{1}{\meter\per\second}), the inertia effects and elasticity are dominating the interfacial instability dynamics \cite{comninou1978can}. An investigation of motions of such natures will reveal new physics unique to the high speed sliding processes, and help elucidate the real friction mechanism in practical application scenarios involving high rate of interfacial separation, including contact dynamics of footwear, tyres, rubbers seals, and brakes of automobiles. 

Therefore, we report in this letter the observations of surface waves in the form of propagating separation zones when elastomer is sliding at high speed on a rigid substrate.
Using total internal reflection imaging technique, we captured the time-resolved morphology of fast-traveling wrinkles at the bottom of elastomer with high-speed recording.
We found that at the high sliding velocities of a few meters per second, waves of detachment move at intersonic velocities, exceeding the Rayleigh and shear wave velocities and approaching the longitudinal wave velocity of the elastomeric material. These results were further validated with numerical simulations under same conditions with high consistency. The spatio-temporal analysis of a simplified plane-strain condition revealed the three different stages of the development of surface wrinkles. It was shown that during the generation, propagation and rebound stage, the detachment initially attached to the moving substrate before its release. The wave velocity continuously increased during the propagation, traversing the Rayleigh/shear wave velocity line and entering the intersonic region. The rebound wave traveled backwards at higher velocity than the forward-propagating waves. Additionally, we discovered both experimentally and numerically the scaling of velocities of detachment waves as stiffness increases. The square-root dependence on stiffness follows the same trends as the other characteristic wave velocities of material. Further analysis revealed that the generation frequencies of these waves conform to the same scaling law. These results suggest an increasing frictional force at elastomer-plastic interface as a function of the stiffness of the bulk material.
On the other hand, as the frequencies of these wrinkles fall within the sensitive hearing range of humans, these discoveries may inspire future studies on the acoustic significance and squeaking mechanisms of rubbery materials, which would guide the rational design of materials for industrial applications involving rubber contacts.
It is important to note that another type of detachment waves, historically named as Schallamach waves \cite{schallamach1971does}, although bearing similar features, have very unique physical origins that distinguish them from the waves discussed in this paper. Schallamach waves are not only much slower in speed, but their origination appears to be governed by interfacial adhesion and relaxation effects \cite{comninou1978can}, whereas elasticity and inertia are dominating the dynamics of the high-speed traveling detachment waves. 

\section{Experimental and Numerical Configurations}

	\subsection{Experimental setup}

We designed an experimental setup to observe fast detachment waves at several meters per second.
Elastomer samples (\SI{40}{\milli\meter} in diameter) are glued at the tip of a ``pendulum'', which consists of a 0.63\,m long arm driven by gravity as shown in Fig.~1(a).
By changing the arm length of the pendulum, the samples are brought to contact with a glass plate at different levels of compression. In our experiments, the local pressures are ranging between \SI{10}{\kilo\pascal} to \SI{40}{\kilo\pascal}.
The arm of the pendulum is heavy enough that the loss of velocity due to the frictional dissipation is less than 5\%.
In our experiments, we varied the velocity of the elastomer samples between \SI{1}{\metre\per\second} and \SI{8}{\metre\per\second}. 

\begin{figure}[ht]
\centering
\includegraphics[width=0.48\textwidth]{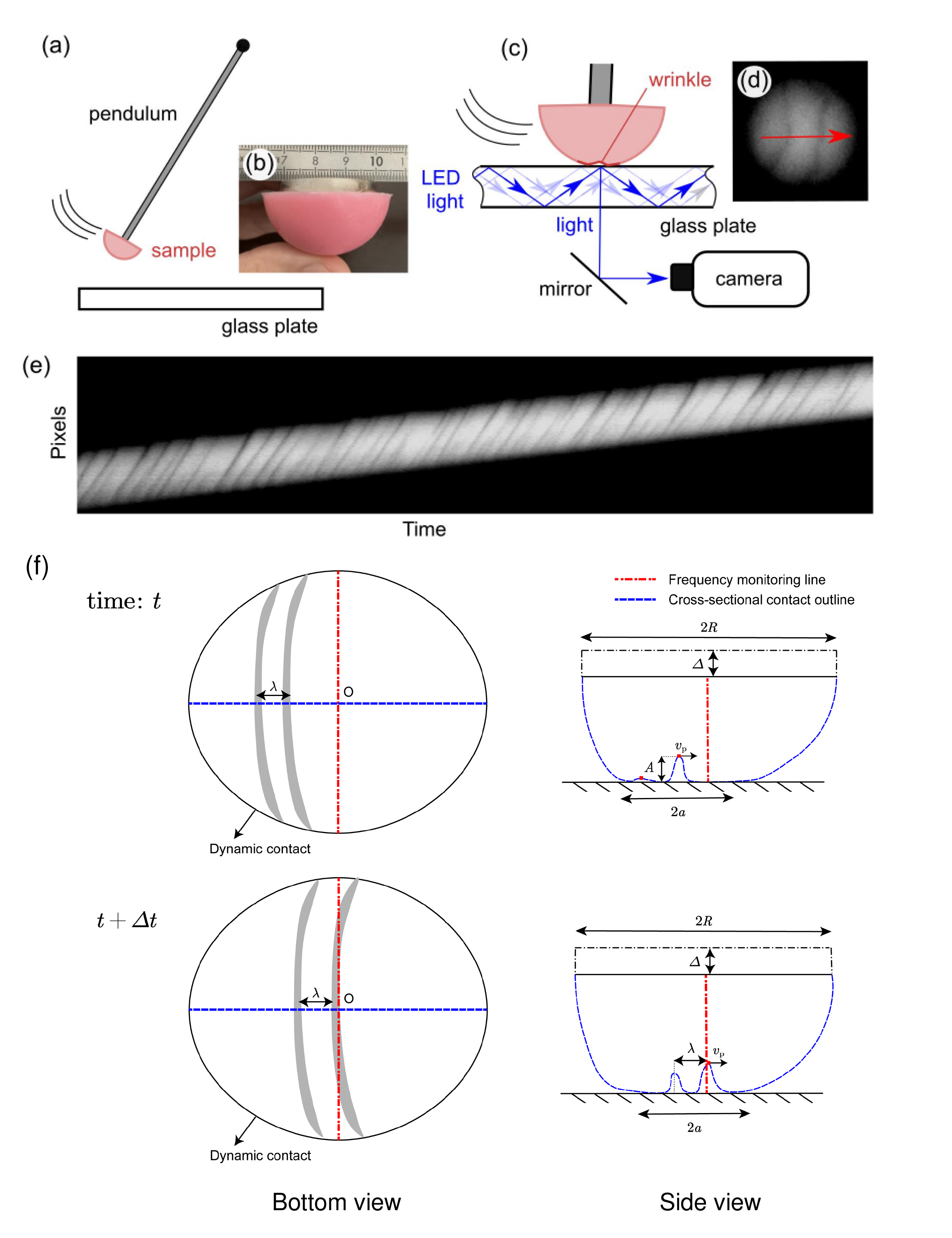}
\caption{Schematics of the experimental setup and spatio-temporal recordings of the contact area change during sliding.
(a) Samples are glued at the tip of a pendulum.
(b) Snapshots of the samples used in this study.
(c) The interface is imaged by using total internal reflection of light in a glass plate. 
(d) The brightness of the interface corresponds to the real area of contact. Here we observe wrinkles travelling in the direction of motion of the sample (left to right). (e) Pixel brightness on the bottom surface (same as in (d)) at different times are stacking up to track the surface elevation along the center line of sample bottom. The scale bar is 5\,mm. (f) illustrations of the essential characteristics of the detachment waves. Bottom view (left) and side view (right) show the calculation of the phase velocity within the contact region. $\lambda$ is the wavelength of the detachment wave and $v_p$ is the phase velocity. $2a$ measures the extent of the contact region. $R$ is the radius of the undeformed sphere, and $\Delta$ is the indentation depth.
}
\label{fig:expsetup}
\end{figure}

The samples are cast in hemispherical molds with a solution of vinylpolysiloxane from Zhermack, as summarized in Table~1.
The elastic modulus is measured in static compression, by monitoring the real contact area as a function of the normal load, assuming pure Hertz contact \cite{johnson1987contact}. 
\begin{table}[ht!]
\begin{tabular}{llcc}
\hline \hline material & color & E (MPa) & $\rho$ (\si{\kilogram\per\metre\cubed})\\
\hline
 Elite Double 8  & pink & 0.3 & 1150\\
 Elite Double 16 & turquoise & 1.1  & 1150\\
 Elite Double 32 & green & 1.5  & 1150 \\
\hline \hline
\end{tabular}
\caption{Parameters of the samples used in this study. The elastic modulus is denoted by $E$ and the density is $\rho$.}
\label{Tab01}
\end{table}

The contact area between the sample and the glass plate is measured by using total internal reflection (TIR) \cite{rubinstein2004detachment, bennett2017contact, dillavou2018nonmonotonic}. 
We used incident LED light below the critical angle for TIR, allowing light to escape only through points of contact, as depicted in Fig.~1(c). 
The brightness of the interface corresponds to the real contact area and allows direct observation of the frictional instabilities, as shown in Fig.~1(d).
The interface is imaged with a high speed camera Phantom v2511 \num{256 x 320} pixels at 100\,kfps and a manual zoom lens Computar 10x.
This provides both an accurate temporal and spatial resolution, respectively 1\,ms and 0.2\,mm. We also carried out a similar spatio-temporal analysis of the simulated sliding contact between an elastomer and a plane surface. 



	\subsection{Finite Element Analysis Setups}
\begin{figure}[h]
		\centering
		\includegraphics[width=0.5\textwidth]{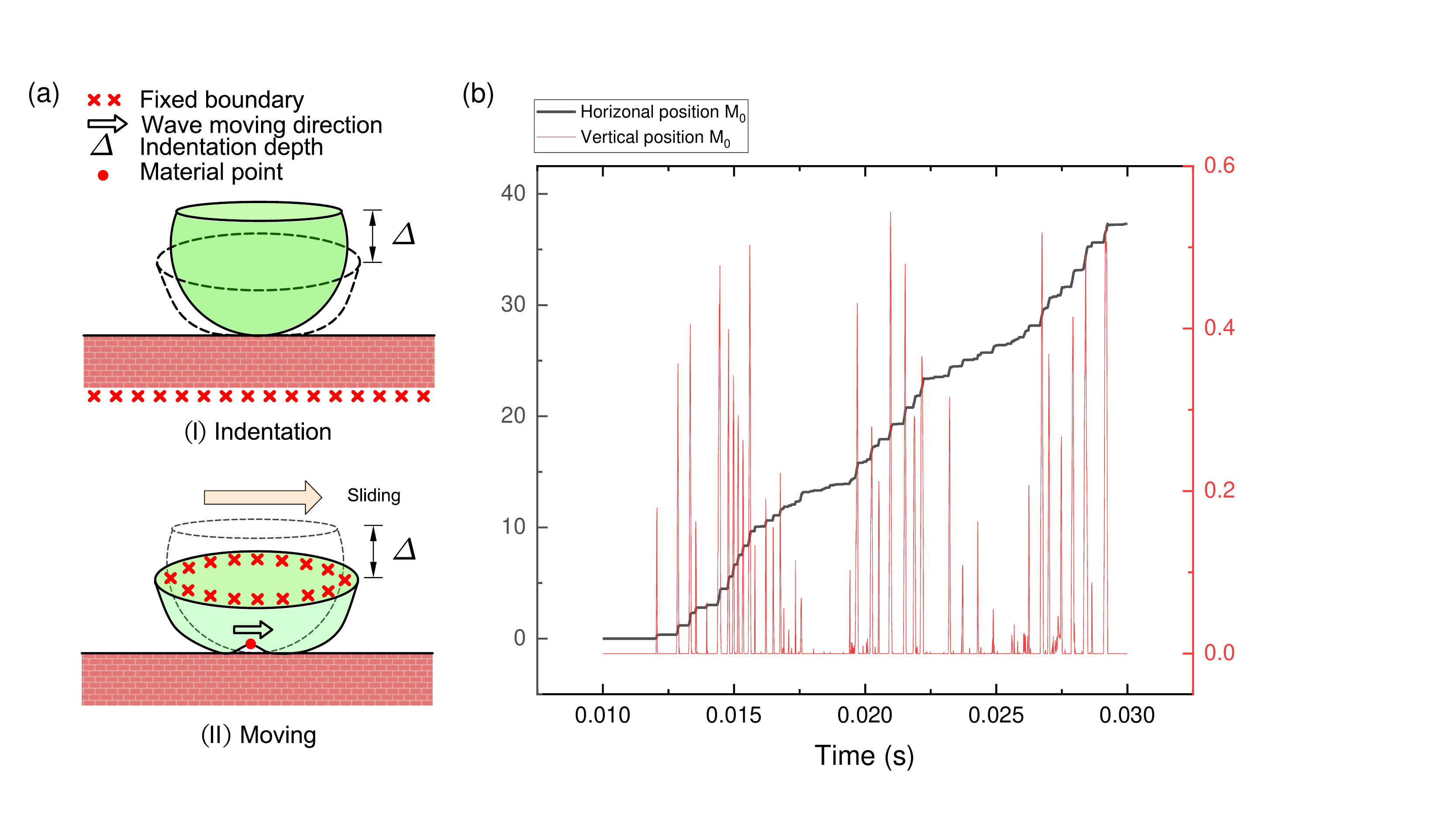}
		\caption{Schematic of the numerical setups and stick-slip plot of the bottom point. (a) The two sequential steps start with (I) indentation of the elastomer against fixated substrate, and proceed to (II) moving of the substrate at constant velocity with elastomer being anchored at top. Friction generated at the interface induces the lift-up and buckling of elastomer surface. (b) Coordinate positions of a material point on the bottom of the elastomer initially at the center of contact. Every horizontal relative movement (to the underlying substrate) of the point is accompanied by a vertical detachment, while no actual slip is observed when the point is in contact.
		}
		\label{fig:numsetup}
	\end{figure}

	The numerical analysis was configured to approach the experimental conditions. A Neo-Hookean constitutive model was adopted to simulate the deformation of elastomer with material constants acquired from \textit{in situ} measurements of different elastomers tested during experiments. Since not only the buckling condition but the wave dynamics of wrinkles were concerned, a dynamic explicit scheme ensured all the necessary information (i.e. wave velocities, frequencies) can be extracted in the time domain.  The underlying substrate was assumed to be rigid. As discussed in the previous section, tangential stress gradients are required for the generation of interfacial instabilities,  therefore both the static and kinetic friction coefficients were used to specify the contact condition at the elastomer-substrate interface. Aiming at capturing the essentials of the wrinkle generation mechanism, the Coulomb type of friction simplifies the loading dependence of the frictional force and we shall go back to reexamine this simplification later.  
	Schematics of the numerical configurations were displayed in Figure \ref{fig:numsetup}. A material point was highlighted at the bottom of the elastomer. Initially situated at the center of the static contact, this point was monitored to keep track of its displacement and to elucidate how the material slides on the plane surface. A correlation plot of the vertical and tangential displacements of the point suggested that in this case, there is no actual relative movement between the two surfaces in contact. The material sliding is solely achieved by the passage of these detachment waves. This echoes with the original observation of Schallamach waves where no detectable slip was recorded between the waves \cite{schallamach1971does}.
	
\section{Characterizations of Fast Detachment Waves}


\subsection{Velocity of the detachment waves}
\begin{figure}[h!]
\centering
\includegraphics[width=0.4\textwidth]{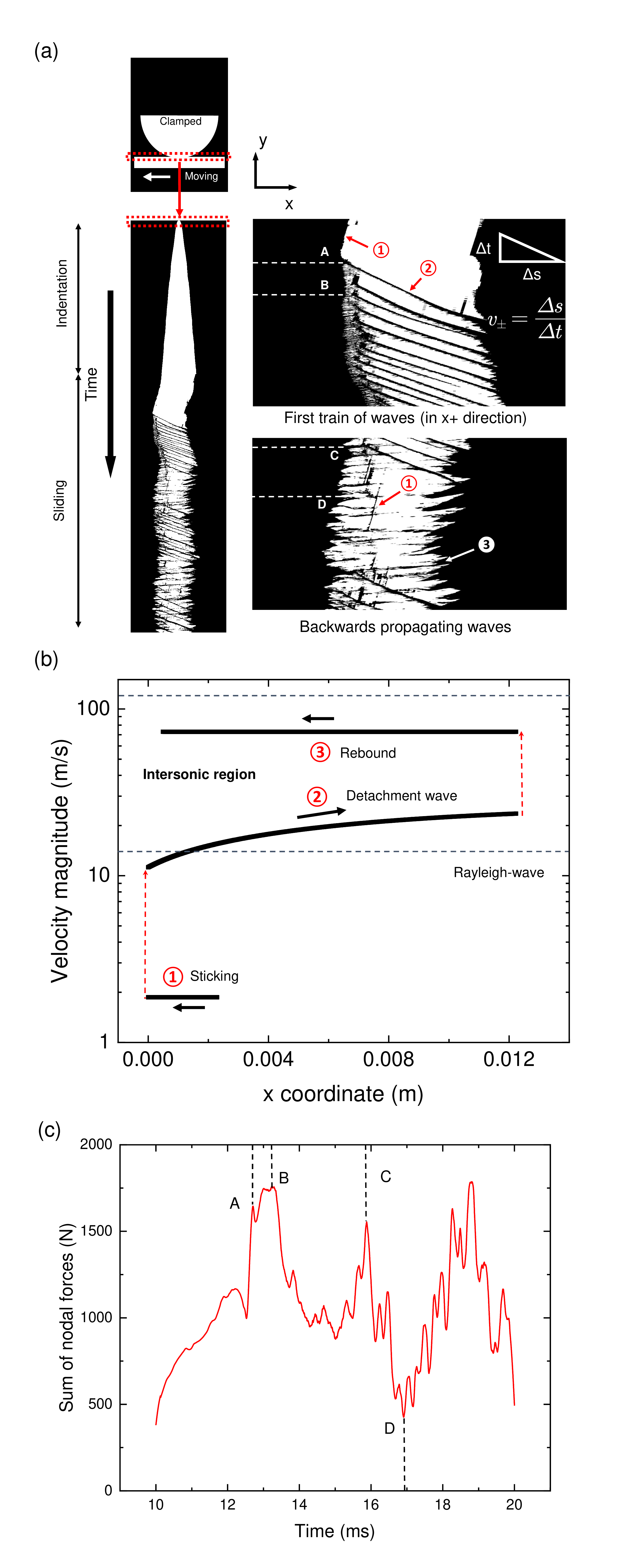}
\caption{(a) Spatio-temporal image of the contact area change during sliding in simulation. Identifiers for three detachment front: \textcircled{1} backward moving contact lines during which elastomer sticks to the substrate and builds up energy. \textcircled{2} forward propagating detachment waves to release the stored energy, and \textcircled{3} is the reflected waves of \textcircled{2} and travels backwards. 
(b) The velocity magnitudes of the detachment. 
(c) The summation of the tangential nodal forces among all bottom nodes in contact. Points A,B,C,D mark different points on Fig. \ref{fig:vel}(a).
}
\label{fig:vel}
\end{figure}

To access the full-field contact information, we first considered numerically a two-dimensional contact where plane-strain condition was assumed with otherwise the same configuration as in Fig. \ref{fig:numsetup}. 
At each frame, the contact area between elastomer and substrate was extracted and different frames were stacked up vertically to constitute the spatio-temporal plot as shown in Figure \ref{fig:vel}(a). During the indentation step, the contact area gradually increased symmetrically with respect to the central line. 
In the following step, as the underlying substrate started to move sideways, the bright area was also distorted because the bottom surface was adhering to the moving substrate with no slipping. The edge of elastomer continued to build up energy (with identifier \circled{1} in Fig.\ref{fig:vel}) until a wrinkle was generated and propagated forward to reach the opposite ( identified as \circled{2}), where waves rebounded and traveled backwards at much higher velocities (notation \circled{3}). The overall contact region also "slipped" back to the original position facilitated by the train of waves, as we could tell from the zoomed-in view of Fig.\ref{fig:vel}(a) that the envelope of first train of waves swung back to the middle line after having being warped to the left by sticking to the substrate. It continued to oscillate around the fixed position and repeated the sticking and snapping-back cycle, and released multiple trains of detachment waves to facilitate the relative movement between elastomer and ground. 

From a more quantitative perspective, on a spatio-temporal plot, the local inclination of lines gives the instantaneous velocity of the wrinkle front, as illustrated in the margin of Fig. \ref{fig:vel}(a). 
We tracked the velocity change of wrinkles in different stages of their development and presented them on the same plot with offset for visualization purpose, since the rebound waves traveled beyond the edge of contact and were not recorded on the plot until they reentered the contact region. The results were shown in Figure \ref{fig:vel}(b) where velocities were extracted by performing polynomial fitting of three different wave signatures in Figure \ref{fig:vel}(a). The first stage followed a constant velocity with value close to that of the moving substrate which the material was sticking to. After storing enough energy, the interfacial wrinkle snapped back and traveled in the direction opposite to the moving ground, initially at speed lower than Rayleigh-wave before crossing the dashed line and entering the intersonic region. The wave velocity monotonically increased until it traveled past the contact edge. The rebound waves propagated at higher and near-constant speed, potentially due to the fact that stress was more uniformly distributed at the interface when reflected wave passed through. 
The emergence of different detachment fronts were closely related to the change in tangential stress as well. As shown in Figure \ref{fig:vel}(c), the tangential stress continued to build up prior to point A where the first detachment started to move. During the A-B section, stress began to plateau and dropped as the propagating waves \textcircled{2} served as a stress release mechanism. The structure gradually returned to its equilibrium position as stress continued to drop between C-D, until new surface detachment \textcircled{1} was generated and the energy build-up process resumed.

	\subsection{Material Properties and Wave Characteristics during Sliding}
\begin{figure}[h!]
\centering
\includegraphics[width=0.35\textwidth]{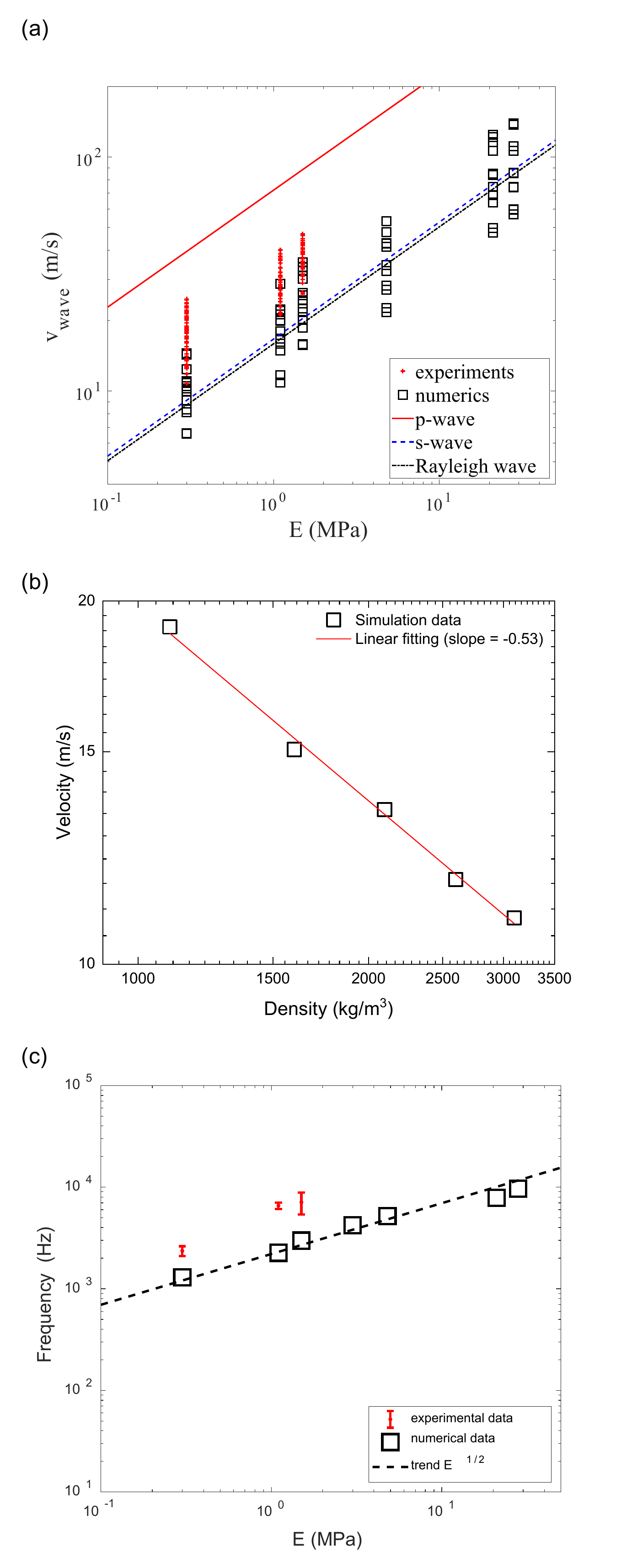}
\caption{(a) Change of the wave velocities as material stiffness increases. Squares indicate simulation data while crosses represent experimental measurements.
(b) Change of the wave velocity as density increases and elastic moduli are fixed.
(c) The frequency readouts from both experiments and simulations as monitored by the "scratch-line" method and plotted against increasing stiffness.  }
\label{fig:freq}
\end{figure}
To verify that the dynamics of these fast detachment waves are indeed dominated by elasticity and inertia, we performed an experimental study in parallel to the simulation of different material properties, and how the wave characteristics were affected by them.
As the realization of plane-strain state is experimentally challenging for elastomer to form a straight and spatially uniform wrinkle, we compared the experimental measurements of detachment wave velocities propagating forward against three-dimensional numerical simulations.
Experimentally, we observed the generation of detachment waves for velocities of samples of a few meters per second.
We performed sets of experiments for three different elastomers spanning a wide range of Young modulus, as reported in Table~1. 
Here we report 144 values of detachment velocities.
Before each measurement, the glass plate was cleaned with a dry wipe and ethanol to remove potential residues.
The detachment wave velocity was systematically observed between the Rayleigh wave velocity and the P-wave velocity, as shown in Fig.~\ref{fig:freq}(a).
Interestingly, the measured velocities follow a simple scaling law, increasing as the square root of the Young modulus of the material, suggesting that the detachment is essentially driven by the elasticity of the samples.
The Young's modulus values that we report in Table~1 and in the x-axis of Fig.~\ref{fig:freq}(a) have been measured in quasi-static compression, and are therefore underestimated values \cite{read1981measurement, jambon2020deformation}. 

We further study the scaling relationship of the detachment wave velocity by performing numerical simulations under the same conditions as the experimental ones, but adding three more materials with $E=$ 4.8, 21 and 28 \SI{}{\mega \pascal}. 
These additional values are typical of shoe materials, which is relevant for practical applications.
If the friction is large enough, elastomer is ``glued'' to the underlying ground and move along with it during the stick phase, up to the point where the maximum static frictional force is reached. 
It then snaps back into its original position during the apparent slip phase with nonzero relative movement occurring at interface. 
During the stick phase, the build-up of frictional stress buckled the elastomer surface which formed a tunnel of air (fluid not directly simulated) in the vicinity of locations with large tangential stress gradients. 
This surface elevation gradually accelerated and moved along the bottom of elastomer from front to trailing end, accompanied by the slip motion of the elastomer and may help release some of the stress. 
Before that wrinkles are generated or within the area between two consecutive waves, elastomer bottom is in complete contact with the substrate. Therefore, phase velocities of detachment waves can be extracted via tracking the locations of abrupt surface elevation, and results from different elastomeric materials were summarized in Fig.~\ref{fig:freq}(a).
The spreading of wave velocities for a single material were attributed to the accelerating nature of these waves, from the generation location up to the point of departing contact edge. 
	
As the stiffness increases, the scaling of detachment wave speeds follow the same relationship as in the experiments, which is also the same scaling as the Rayleigh, shear and longitudinal waves of elastomer, as shown in Fig.~\ref{fig:freq}(a). 
Previously, interfacial waves were considered to be much slower than the characteristic wave speeds of the material, typically in the mm or $\mu$m ranges. 
The results from our observations and simulations suggest the velocities of Schallamach waves can well exceed the shear wave velocity, entering the intersonic region of the material. As a side note, we also varied the densities of the materials while keeping the elastic constants unchanged. The linearly-fitted slope of -0.53 in Fig.~\ref{fig:freq}(b) was consistent with Fig.~\ref{fig:freq}(a), suggesting the combined effects of elasticity and inertia control the dynamics of these detachment waves.

To find the correlation between frequency of waves and stiffness, we specified a single-element width line across the bottom of spherical elastomer transverse to the moving direction of ground. The accumulative contact area of all the elements on this line was monitored in real-time for the duration of the simulation. Our analysis showed the frequency of surface detachment passing through the monitoring line observes the same scaling law as the velocity of waves when stiffness varies, as shown in Figure \ref{fig:freq}(c). We hence provide a non-dimensional analysis of this result. The momentum balance equation in the x-direction dictates that $\rho \frac{\partial ^2u}{\partial t^2}\propto \frac{\partial \sigma _{yx}}{\partial y} $. Between successive waves, We estimated the variation in displacement to be $\lambda $ on the same order as the wavelength (see Figure \ref{fig:expsetup}(f) ), and the time increment to be $ 1/f $. The tangential stress can be approximated as $G\epsilon_{yx} \propto G\frac{\lambda}{2a} $ following a similar argument as in \cite{rand2006insight}, and the increment in y-direction as a characteristic length along the surface normal ($A$). The contact region $2a$ can be estimated with $R$ and $\Delta$ as $a\sim\sqrt{R\Delta}$. Therefore, we have
\begin{align*}
    \rho \lambda f^2 &\propto G \frac{\lambda}{2aA} \sim G \frac{\lambda}{2\sqrt{R\Delta}A} \\
    \Rightarrow f &\propto \sqrt{\frac{G}{\rho}}
\end{align*}
This simplified analysis did not take into account the viscoelastic effects, and they would likely to affect the wavelength at a higher frequency. Overall, an increase in the material stiffness leads to a faster relaxation process, so we would expect the frequency of wave generation to grow as well, which was plotted in Fig.\ref{fig:freq}(b) for three different materials tested in the experiments. It has important physical implications and acoustic significance, because at the macroscopic sliding velocity (on the order of \SI{1}{\metre\per\second} )and application-relevant material stiffness (e.g. \SI{10}{\mega\pascal} is close to the stiffness of rubber shoes) the frequency of instabilities not only overlap with human hearing range and produce sensational discomfort, but could also serve as indications of high material abrasion rate and poorly lubricated connections which may involve safety concerns as well. The understanding of the interaction between the dynamical instability and surrounding medium at the frictional contact surface of soft material will help elucidate the noise generation mechanism and inspire solutions to mitigate the detrimental effects, which will be the focus of our future work along this line.

In conclusions, we have reported on our observations that surface detachment waves 
exist in the intersonic region of the elastomeric material. To the best of our knowledge, it is also the first time that phase velocities of these detachment waves had been experimentally verified to exceed shear wave velocity of the underlying material, in comparison  to previous studies of Schallamach waves where speeds mainly fall within the sub-millimeter per second regions. An immediate consequence of these macroscopic velocities is that since the wave generation frequency, alongside the wave velocity, scales with square root of stiffness of material, it readily enters the audible frequency range of human beings and can be correlated with other vibratory frequencies which jointly create undesirable and detrimental noises in many daily and industrial applications, colloquially referred to as squeaking, squealing, and screeching of elastomers. 
Our analysis of the wave signatures on the spatio-temporal plot suggested at least three different wave fronts can be extracted and connected to the corresponding development stages of the life cycle of a surface detachment, which provides insights on how these dynamic instabilities are generated, propagated and eventually damped out by material and interfacial dissipation.
Further studies on the interaction between soft materials and ambient fluid in the boundary layer could promote better understandings of acoustic consequences of these intersonic surface waves.

\nocite{*}

\bibliography{apssamp}


\appendix
\counterwithin{figure}{section}
\section{Local stress distribution during the initiating stage of detachment waves}
\begin{figure}[h!]
	\centering
	\includegraphics[width=0.5\textwidth]{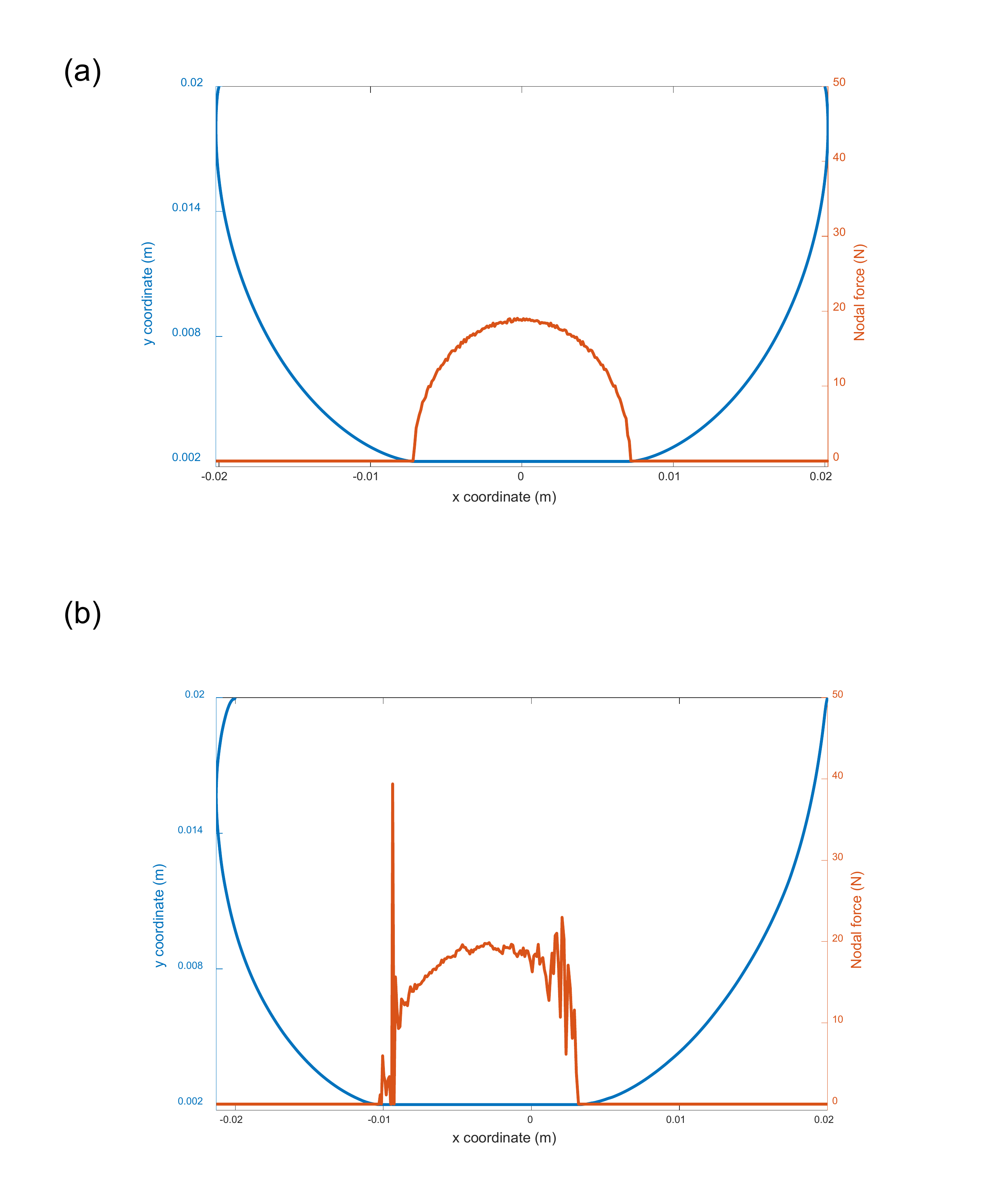}
	\caption{(a) Distribution of the normal nodal force along the bottom surface when compression is complete and before the sliding begins, and (b) after sliding and right before the first detachment wave is generated. The normal stress distribution can be seen to change from being symmetrical to biased where there is large stress gradient near the detachment generation point.}
	\label{fig:appsigma}
\end{figure}

The stress distribution along the bottom of the elastomer was extracted and illustrated in Figure \ref{fig:appsigma}. At the end of normal indentation, there was no tangential displacement and stress followed a symmetrical distribution as in Figure \ref{fig:appsigma}(a). When sliding was initiated and right before the first detachment wave was about to be generated near the left edge of contact, there was a huge spike in the stress distribution since large stress gradient was required to generate a surface detachment.

\end{document}